\documentclass[%
 reprint,
 amsmath,amssymb,
 aps,prl,
]{revtex4-2}

\usepackage{graphicx}
\usepackage{dcolumn}
\usepackage{bm}
\usepackage{color}

\begin{document}

\preprint{APS/123-QED}

\title{Glass Transition in Monolayers of Rough Colloidal Ellipsoids}%
\author{Jian Liang$^{1}$}
\thanks{These authors contributed equally.}
\author{Xuan Feng$^2$}
\thanks{These authors contributed equally.}
\author{Ning Zheng$^3$}
\author{Huaguang Wang$^{1}$}
\email{hgwang@suda.edu.cn}
\author{Ran Ni$^2$}
\email{r.ni@ntu.edu.sg}
\author{Zexin Zhang$^{1,4}$}
\email{zhangzx@suda.edu.cn}
\affiliation{$^{1}$College of Chemistry, Chemical Engineering and Materials Science, Soochow University, Suzhou 215123, China}
\affiliation{$^{2}$School of Chemistry, Chemical Engineering and Biotechnology, Nanyang Technological University, 62 Nanyang Drive, 637459, Singapore}
\affiliation{$^{3}$School of Physics, Beijing Institute of Technology, Beijing 100081, China}
\affiliation{$^{4}$Institute for Advanced Study, Center for Soft Condensed Matter Physics and Interdisciplinary Research, School of Physical Science and Technology, Soochow University, Suzhou 215006, China}


\begin{abstract}
Structure-dynamics correlation is one of the major ongoing debates in the glass transition, although a number of structural features have been found connected to the dynamic heterogeneity in different glass-forming colloidal systems. Here using colloidal experiments combined with coarse-grained molecular dynamics simulations, we investigate the glass transition in monolayers of rough colloidal ellipsoids. Compared with smooth colloidal ellipsoids, the surface roughness of ellipsoids is found to significantly change the nature of glass transition. In particular, we find that the surface roughness induced by coating only a few small hemispheres on the ellipsoids can eliminate the existence of orientational glass and the two-step glass transition found in monolayers of smooth ellipsoids. This is due to the surface roughness-induced coupling between the translational and rotational degrees of freedom in colloidal ellipsoids, which also destroys the structure-dynamics correlation found in glass-forming suspensions of colloidal ellipsoids. Our results not only suggest a new way of using surface roughness to manipulate the glass transition in colloidal systems, but also highlight the importance of detailed particle shape on the glass transition and structure-dynamics correlation in suspensions of anisotropic colloids.
\end{abstract}

\maketitle


As the Nobel Prize-winning physicist Philip Warren Anderson wrote in 1995, ``The deepest and most interesting unsolved problem in condensed-matter physics is probably the theory of the nature of the glassy state and the glass transition"~\cite{Anderson1995}, and one of the major ongoing debates is the structure-dynamics correlation in the glass transition~\cite{Tanaka2019}. 
Because of the controllability over physical and chemical properties of particles, like shape and interactions~\cite{Eckert2002, Fullerton2020, Kramb2010, Zheng2011, Mishra2013,kang2013prl, Liu2020}, colloids have been employed as excellent model systems for investigating the physics of glass transition in recent decades. In glass forming colloidal hard spheres, it has been found that the glassy dynamics can be related to locally favored structures formed in the system~\cite{Tong2019, Zhang2024}, e.g., particles of crystal-like~\cite{Kawasaki2007, Tanaka2010}, icosahedral~\cite{Tanaka2012}, or tetrahedral order~\cite{Frank2020} or some machine learned structural features~\cite{Boattini2020}. In glass-forming anisotropic colloids, the structure-dynamics correlation was found to be more pronounced. For instance, in monolayers of colloidal ellipsoids, it was observed that with increasing packing density, clusters of orientationally aligned particles appear, in which the rotation of orientationally aligned ellipsoids slows down significantly. This leads to the two-step glass transition, a translational glass, and an orientational glass where the rotation of particles is glassy while the translation remains fluid-like~\cite{Zheng2011, Zheng2014, Zheng2021}. This orientational glass was subsequently also found in systems of colloidal ellipsoids with short-range attraction~\cite{Mishra2013, Mishra2014} as well as colloidal ellipsoids in 3D~\cite{Roller2021}.

Although the surface roughness, a crucial physical property of particles, is known to significantly influence particle dynamics and rheology in colloidal systems~\cite{Hsu2021, Hsiao2017, Hus2018, Ilhan2020, Kato2023, Lu2023}, only recently the glass transition of rough colloidal spheres was studied.  It was found that the surface roughness induces anomalous rotational dynamics capable of altering the nature of glass transition \cite{Ilhan2022, Ilhan2020}, underscoring the critical role of particle roughness in the glass transition. However, the effect of surface roughness on the glass transition of anisotropic colloids,  especially the strong structure-dynamics correlation found, remains unknown.

In this Letter, using both colloidal experiments and coarse-grained molecular dynamics (MD) simulations, we investigate the glass transition in monolayers of rough colloidal ellipsoids. These rough ellipsoids are fabricated by coating small hemispheres randomly on the ellipsoid surfaces. We find that introducing even a tiny amount of surface roughness can significantly change the nature of glass transition. In particular, the glass transition becomes one-step, which is in marked contrast to the two-step glass transition in smooth ellipsoids of the same aspect ratio. This is due to the fact that the surface roughness induces strong coupling between the translation and rotation of the rough ellipsoids, thus eliminating the structure-dynamics correlation and the corresponding two-step glass transition previously found in smooth ellipsoids.

\begin{figure}
\centering
\includegraphics[width=\linewidth]{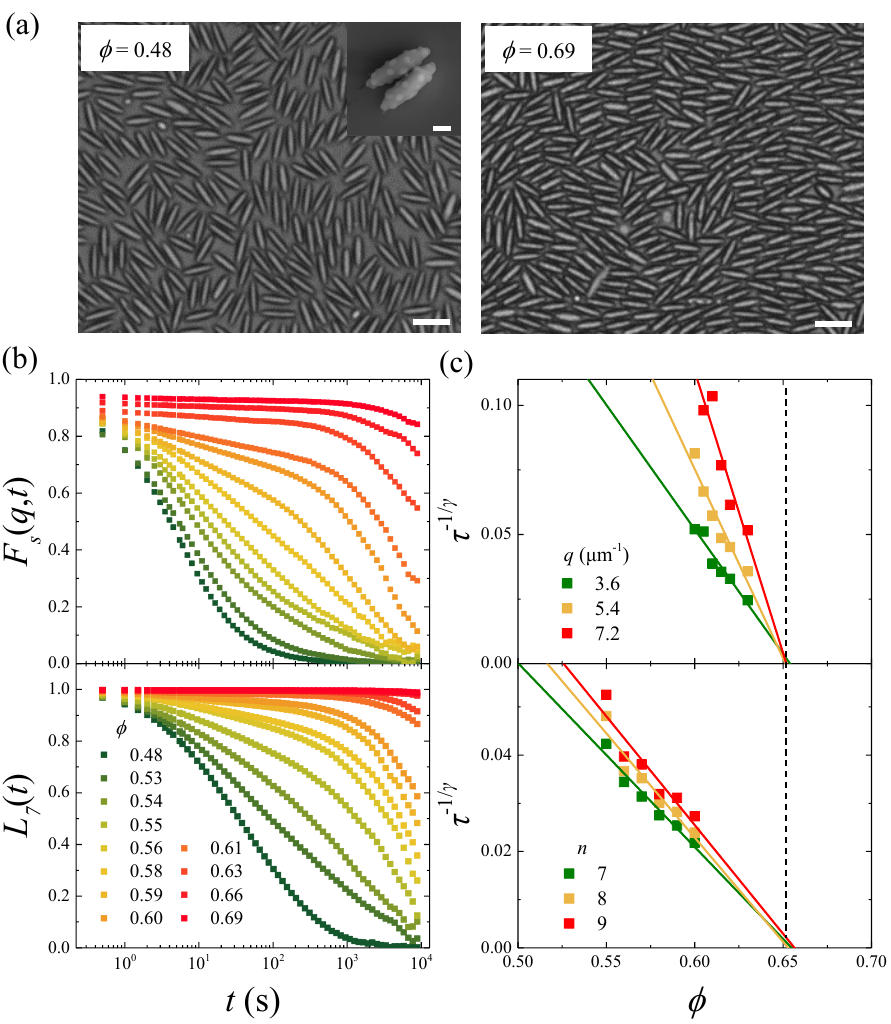}
\caption{
(a) Brightfield micrographs of monolayers of rough ellipsoids, in which only part of the field of view is shown for clarity. Scale bars: 10 $\mu$m. Inset: SEM image of rough ellipsoids. Scale bar: 2 $\mu$m. (b) $F_s(q,t)$ (upper panel) and $L_7(t)$ (bottom panel) at different $\phi$. $q = 3.6 \mu {\rm m}^{-1}$, taken as the first peak position in the structure factor. (c) The fitted relaxation time $\tau(\phi)\sim(\phi_g-\phi)^{-\gamma}$. $\gamma = 2.64\pm 0.01$ and $2.59\pm 0.01$ for translational (upper panel) and rotational (lower panel) relaxation times, respectively. The scalings suggest a one-step glass transition, marked by the vertical dashed line. }
\label{fig1}
\end{figure}

\begin{figure}
\centering
\includegraphics[width=\linewidth]{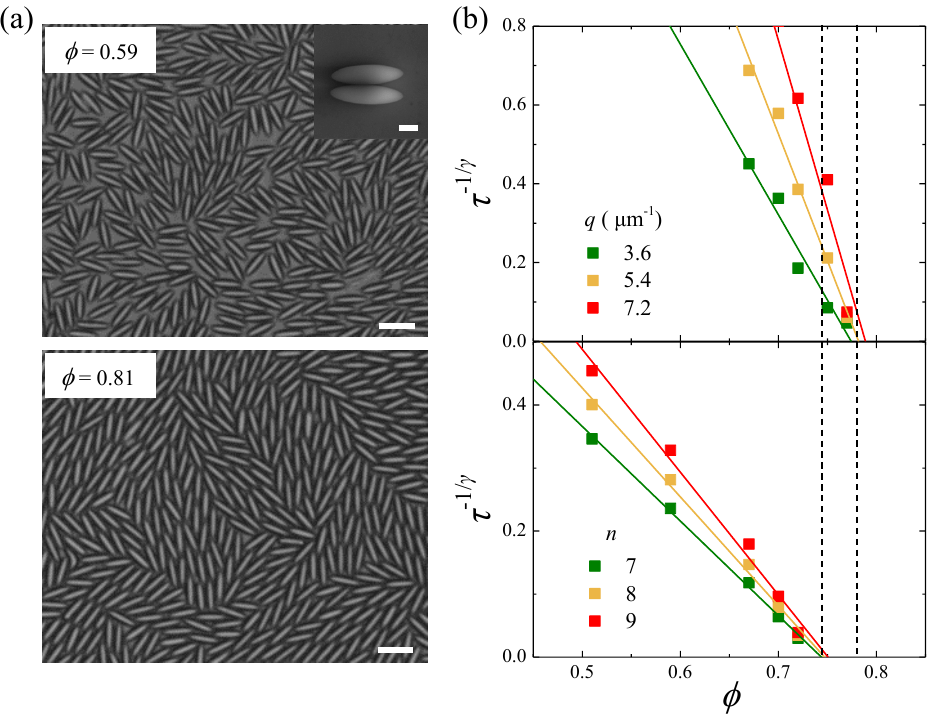}
\caption{
(a) Brightfield micrographs of monolayers of smooth ellipsoids, in which only part of the field of view is shown for clarity. Scale bars: 10 $\mu$m. Inset: SEM image of smooth ellipsoids. Scale bar: 2 $\mu$m. (b) The fitted relaxation time $\tau(\phi)\sim(\phi_g-\phi)^{-\gamma}$. $\gamma = 2.71\pm 0.01$ and $2.36\pm 0.01$ for translational (upper panel) and orientational (lower panel) relaxation times, respectively. The scalings suggest a two-step glass transition, marked by the two vertical dashed lines.}
\label{fig2}
\end{figure}

The rough ellipsoids were synthesized by stretching pre-fabricated polystyrene spheres \cite{Ho1993}, where small polysiloxane colloids grew to generate rough surfaces (inset of Fig. 1(a) and Fig. S1). The obtained ellipsoids had a small polydispersity of 8\% with an aspect ratio of 3.8, which is the ratio between the long and short axes of the ellipsoid. The diameter of the short axis is $\sigma \approx 2.0 \mu{\rm m}$, and each ellipsoid is coated with $n_s \approx 20$ hemispheres of diameter $\sigma_s$ ranging from 400 to 1000 nm. A similar stretching method was applied to prepare smooth ellipsoids with the same aspect ratio by using smooth polystyrene spheres (Fig. S1). The cleaned ellipsoids were suspended in an aqueous solution of 2 mM sodium dodecyl sulfate, and the ionic strength in the suspension resulted in a small Debye length of the ellipsoids making them moderately hard particles \cite{Zheng2011, Liu2020}. The suspension of particles was confined between two glass coverslips producing colloidal monolayers (Fig. 1(a) and Fig. 2(a)). To study the glass transition, monolayers at different particle areal fractions ($\phi$) are investigated; $\phi = A\rho$, where $A$ is the cross-sectional area of the particle and $\rho$ is the number density averaged over all video frames.
Experimental details are included in~\cite{supinfo}.

\begin{figure*}
\centering
\includegraphics[width=0.9\linewidth]{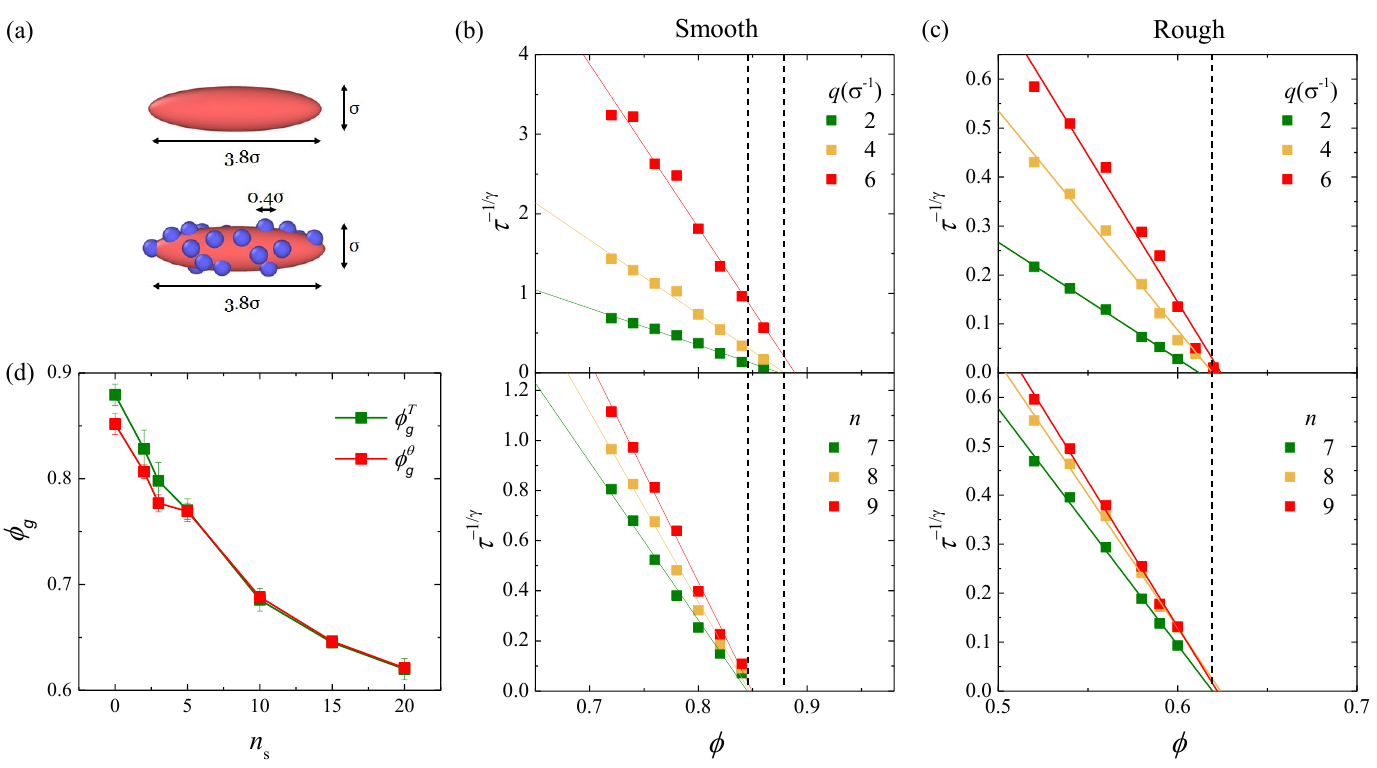}
\caption{
(a) Schematic of smooth and rough ellipsoids in simulation. (b,c) The fitted relaxation time obtained from simulations $\tau(\phi)\sim(\phi_g-\phi)^{-\gamma}$ for the smooth ellipsoid (b) and the rough ellipsoid (with $n_s = 20$ spheres on each ellipsoid) (c). In (b), $\gamma = 1.49\pm 0.38$ (upper panel) and $3.01\pm 0.02$ (lower panel), and the scalings suggest a two-step glass transition, marked by the two vertical dashed lines. In (c), $\gamma = 1.91\pm 0.13$ (upper panel) and $3.65\pm 0.07$ (lower panel), and the scalings suggest a one-step glass transition, marked by the vertical dashed line. The smallest value of $q$ in (b,c) is taken as the first peak position in the structure factor. (d) Glass transition density $\phi_g^{T}$ and $\phi_g^{\theta}$ in monolayers of rough ellipsoids as a function of $n_s$ the number of smooth spheres on each ellipsoid.
}
\label{fig3}
\end{figure*}

The translational and rotational relaxations are characterized by the self-intermediate scattering function $F_s(q,t) \equiv \frac{1}{N}\langle \sum^{N}_{j=1}e^{i\mathbf{q}\cdot[\mathbf{r}_j (t)-\mathbf{r}_j (0)]} \rangle$ and orientation correlation function $L_n(t) \equiv \frac{1}{N}\langle\sum^{N}_{j=1}$cos$n[\theta_j(t)-\theta_j(0)] \rangle$, respectively, where $\mathbf{r}_j (t)$ and $\theta_j(t)$ are the position and orientation of particle \emph{j} at time \emph{t} with \emph{N} the total number of particles, $\mathbf{q}$ the scattering vector, \emph{n} a positive integer, and $\langle \cdot \rangle$ calculates the ensemble average.
Figure 1(b) plots the time evolution of $F_s(q,t)$ and $L_n(t)$ in monolayers of rough ellipsoids at various $\phi$. At low $\phi$ both $F_s(q,t)$ and $L_n(t)$ develop one-step relaxations, implying a fluid phase. At high $\phi$ both functions experience two-stage relaxations corresponding to motion within cages of neighboring particles at short times and cage-breaking structural rearrangements at long times, representing the typical glassy dynamics. The change of $F_s(q,t)$ and $L_n(t)$ with increasing $\phi$ for rough ellipsoids are similar to that for smooth ellipsoids (Fig. S2), a hallmark of the classic relaxation dynamics during the glass transition.

Mode-coupling theory (MCT) scaling of relaxation time is applied to determine the glass transition points~\cite{Zheng2011, Zheng2014, Mishra2013}, and the relaxation time $\tau$ is defined as the time when $F_s(q,t)$ and $L_n(t)$ decays to $e^{-1}$ \cite{Kob1994, Mishra2013, Liu2020}. According to MCT, the relaxation time $\tau$ diverges algebraically at $\phi_g$: $\tau(\phi)\sim(\phi_g-\phi)^{-\gamma}$, with which we fit the obtained relaxation time $\tau$ for rotation and translation at different $q$ and $n$ as shown in Fig. 1(c). One can see that $\tau^{-1/\gamma}$ linearly depends on $\phi$ for different choices of \emph{q} and \emph{n}, and intriguingly, all the fits suggest that the translation and rotation of particles are coupled, and both glass transitions occur at the same particle density, $\phi_g^T = \phi_g^\theta = 0.65$ (Fig. 1(c)). Across this density, both translational and rotational non-Gaussian parameters exhibit no distinct peaks and decrease with time (Fig. S3), and the sizes of clusters of both translational and rotational fast particles decrease, reflected by the narrow probability distributions of the sizes (Fig. S4). This signifies the glass formation for both translational and rotational degrees of freedom \cite{Weeks2000, Zheng2011} confirming a one-step glass transition. For smooth, long ellipsoids, i.e., aspect ratio $> 2.5$, previous work has found that the rotational glass transition occurs at a lower density than the translational glass transition, resulting in a two-step glass transition \cite{Zheng2011, Zheng2014}. Here, for comparison, we also investigate the glass transition for smooth ellipsoids with the same aspect ratio 3.8 as the rough ellipsoids (Fig. 2). The MCT fits show a two-step glass transition for the smooth ellipsoids with $\phi_{g}^T = 0.78 \pm 0.01$ and $\phi_{g}^{\theta} = 0.74 \pm 0.01$, and this agrees with Ref.~\cite{Zheng2011, Zheng2014}. 

\begin{figure*}
\centering
\includegraphics[width=\linewidth]{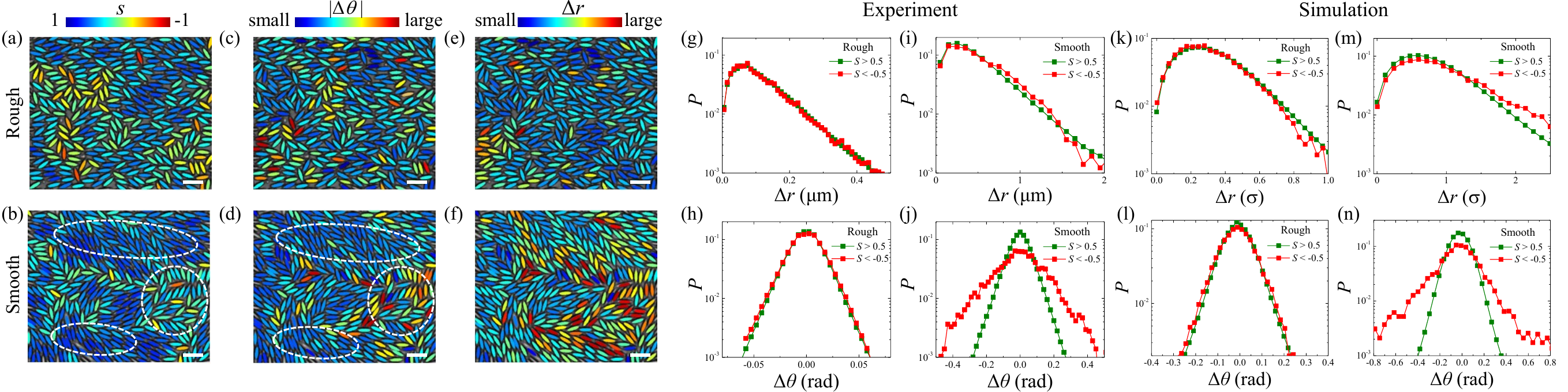}
\caption{
(a-f) Representative spatial distributions of local nematic order parameters $S$ (a,b), rotational displacements (c,d), and translational displacements (e,f) for rough ellipsoids at $\phi = 0.63$ (top panel) and smooth ellipsoids at $\phi = 0.72$(bottom panel) in experiments. Scale bars: 10 $\mu$m. The dashed ellipses in (b) and (d) highlight the particles with large(small) $S$ that exhibit small(large) rotational displacements. (g-j) The corresponding probability distributions of translational displacements, $\Delta r$, and rotational displacements, $\Delta\theta$, for particles with large $S>0.5$ and small $S<-0.5$. (k-n) Simulation results of the probability distributions for rough (at $\phi = 0.61$) and smooth (at $\phi = 0.84$) ellipsoids.}
\label{fig4}
\end{figure*}

These results suggest that the surface roughness can significantly affect the structural relaxation in monolayers of glass-forming ellipsoids. To understand the underlying physics, we perform MD simulations of ellipsoids in 2D~\cite{thompson2022lammps, swope1982computer, nguyen2011rigid, miller2002symplectic, schneider1978molecular}. In MD simulations, the interaction between smooth ellipsoids is modelled by a continuous WCA-like Gay-Berne (GB) potential~\cite{weeks1971role, de1991liquid,berardi1998gay, berardi2008field, brown2009liquid,jayaram2020scalar}, and rough ellipsoids are modelled by coating small hemispheres on the smooth ellipsoids randomly (Fig. 3(a)). In rough ellipsoids, the ellipsoid-hemisphere and hemisphere-hemisphere interactions are also modelled by WCA-like GB potentials, where the diameter of hemispheres is fixed at $\sigma_s = 0.4 \sigma$. Simulation details are included in~\cite{supinfo}. The MCT fits of the relaxation times obtained from MD simulations in monolayers of smooth and rough ($n_s = 20$) ellipsoids are shown in Fig. 3(b) and (c), respectively, where the aspect ratio of ellipsoids is 3.8, same as in experiments. The simulation results suggest a two-step glass transition in smooth ellipsoids (Fig. 3(b)) and a one-step glass transition in rough ellipsoids (Fig. 3(c)), in excellent agreement with experiments. Both our experiments and simulations confirm the transformation from a two-step glass transition to a one-step glass transition for the ellipsoids when introducing roughness on their surfaces. To determine the exact degree of the roughness that induces this transformation, in MD simulation, the roughness is tuned by changing $n_s$ from 0 to 20 to vary the degree of roughness. One can see that with increasing $n_s$, both $\phi_g^T$ and $\phi_g^\theta$ decreases, and their difference also decreases and vanishes when $n_s \ge 5$ (Fig. 3(d)). In MD simulations, ellipsoids are moving in a 2D plane and not allowed to spin or rotate out of the plane. Therefore, coating $n_s = 5$ hemispheres on an ellipsoid is a very small change in roughness, which proves sufficient to eliminate the two-step glass transition and corresponding orientational glass.

Previous studies on monolayers of smooth ellipsoids show that the large aspect ratio of particles can induce local orientational ordering represented by nematic domains~\cite{Zheng2011}. Within these domains, the rotation of ellipsoids gets suppressed as it demands cooperative motions across many particles while the translation remains feasible as ellipsoids can glide readily without extensive collective motions. This leads to the existence of orientational glass~\cite{Zheng2011}. When the aspect ratio drops below 2.5, the nematic domains disappear and the two-step glass transition transforms into a one-step transition~\cite{Mishra2013, Zheng2014}. This suggests that the glassy rotational dynamics in the orientational glass is correlated to the local orientational order, serving as the structural origin of the decoupling of translational and rotational motions and the two-step glass transition. However, for the rough ellipsoids, although the system is capable of forming nematic domains, the rough surfaces of the particles significantly hinder the gliding of particles and hence the translational motions. Therefore, both translational and rotational motions are impeded simultaneously. As a result, orientational glass disappears and the one-step glass transition is observed.

To understand the disappearance of orientational glass in rough ellipsoids, we calculate the local orientational order characterized by the local nematic order parameter $S = \langle \cos (2 \Delta \theta)\rangle$, where $\Delta\theta$ is the angle between the orientation of the particle and its nearest-neighbors and $\langle \cdot \rangle$ denotes an average of all nearest-neighbors. For an ellipsoid, $S = 1$ when it is perfectly locally aligned, and $S = -1$ when the nearest-neighbors are completely not aligned with the particle.  The spatial distributions of $S$ in typical monolayers of rough ($n_s=20$) and smooth ellipsoids near glass transition are plotted. It is found that orientationally aligned clusters appear in both systems reflected by groups of particles with large $S$ (Fig. 4(a, b)). This indicates that the roughness has a negligible effect on the particle orientational order, which is corroborated by the similar increase of the averaged $S$ over systems with increasing $\phi$ for rough and smooth ellipsoids (Fig. S5).

To anatomize the correlation between the local structures and the particle dynamics, in Fig. 4(c-f), we plot the corresponding spatial distributions of translational and rotational displacements over $t^*$ corresponding to the maximum of the non-Gaussian parameter of displacements (Fig. S3)~\cite{Zheng2011}. In smooth ellipsoids, we find a clear correlation between the spatial distribution of rotational displacements (Fig. 4(d)) and the local orientational order (Fig. 4(b)), in which the rotation of highly orientationally aligned particles is significantly suppressed, and this agrees with Ref.~\cite{Zheng2011}.  However, in rough ellipsoids, such structure-dynamics correlation does not exist, and the rotation of orientationally aligned particles is similar to that of other particles.

The correlations between rotational dynamics and local orientational order are further characterized by the structure-dependent probability distributions of the particle motions. We determine the orientationally aligned ($S>0.5$) and disaligned particles ($S< -0.5$) in both rough and smooth ellipsoids, and then plot the probability distribution of translational and rotational displacements of these particles. For translational displacements, the probability distributions are nearly identical for orientationally aligned and disaligned particles in both rough and smooth cases (Fig. 4(g, i)). For rotational displacements, however, the probability distribution for orientationally disaligned particles is much wider than that of orientationally aligned particles in smooth ellipsoids (Fig. 4(j)), signifying the greater rotational mobility of orientationally disaligned particles ~\cite{Mishra2013, Liu2020, Zheng2021}. By contrast, no difference between the distributions of orientationally aligned and disaligned particles is observed in rough ellipsoids (Fig. 4(h)). All these experimental findings are reproduced in MD simulations (Fig. 4(k-n)). Note that the packing densities for systems of rough and smooth ellipsoids in Fig. 4 are chosen near $\phi_g$ which is generally used to demonstrate the significant correlation between the structure and dynamics~\cite{Tanaka2010, Zheng2021}.

In conclusion, we have combined experiments and MD simulations to study the glass transition in monolayers of rough colloidal ellipsoids. In contrast to the two-step glass transition and strong structure-dynamics correlations observed in monolayers of smooth ellipsoids, rough ellipsoids undergo a one-step glass transition, where both the translational and rotational glass transitions occur at the same particle density. Our MD simulations show that decorating only a few small hemispheres on colloidal ellipsoids can eliminate the two-step glass transition and structure-dynamics correlation in monolayers of glass-forming ellipsoids. Our results question the generality of the structure-dynamics correlations found in smooth glass-forming particles and highlight the importance of detailed particle morphology in the glass transition of anisotropic colloids.




\begin{acknowledgments}
This work was financially supported by the National Natural Science Foundation of China (12074275, 12474196, and 11704269), the Natural Science Foundation of the Jiangsu Higher Education Institutions of China (20KJA150008 and 17KJB140020), the PAPD program of Jiangsu Higher Education Institutions, the State and Local Joint Engineering Laboratory for Novel Functional Polymeric Materials, the Jiangsu Key Laboratory of Advanced Functional Polymer Materials, the Jiangsu Engineering Laboratory of Novel Functional Polymeric Materials, the Academic Research Fund from the Singapore Ministry of Education (RG151/23 and MOE2019-T2-2-010) and the National Research Foundation, Singapore, under its 29th Competitive Research Program (CRP) Call (NRF-CRP29-2022-0002).
\end{acknowledgments}

\bibliography{manuscript}

\end{document}